# Scale effects on spatially embedded contact networks

**Peng Gao, Ling Bian\***

*Department of Geography, University at Buffalo, The State University of New York, Buffalo, NY 14261, United States*

**Abstract**

Spatial phenomena are subject to scale effects, but there are rarely studies addressing such effects on spatially embedded contact networks. There are two types of structure in these networks, network structure and spatial structure. The network structure has been actively studied. The spatial structure of these networks has received attention only in recent years. Certainly little is known whether the two structures respond to each other.

This study examines the scale effects, in terms of spatial extent, on the network structure and the spatial structure of spatially embedded contact networks. Two issues are explored, how the two types of structures change in response to scale changes, and the range of the scale effects. Two sets of areal units, regular grids with 24 different levels of spatial extent and census units of three levels of spatial extent, are used to divide one observed and two reference random networks into multiple scales. Six metrics are used to represent the two structures.

\* Corresponding author at: Department of Geography, University at Buffalo, 120 Wilkeson Quad, Amherst, NY 14261, United States. Tel.: +1 716 645 0484; fax: +1 716 645 2329.

*E-mail addresses*: pgao3@buffalo.edu (P. Gao), lbian@buffalo.edu (L. Bian).

Results show different scale effects. In terms of the network structure, the properties of the observed network are sensitive to scale changes at fine scales. In comparison, the clustered spatial structure of the network is scale independent. The behaviors of the network structure are affected by the spatial structure. This information helps identify vulnerable households and communities to health risks and helps deploy intervention strategies to spatially targeted areas.

**Key words**: Scale effect; Spatially embedded networks

# 1 Introduction

Human contact networks play a critical role in disease dispersion, as repeatedly stressed in reports on some of the most dangerous communicable diseases, such as SARS, Avian Flu (H5N1), and Ebola (Chan 2014; Ferguson et al. 2005; Ferguson et al. 2006; Riley et al. 2003). A 'contact network' refers to a network of human contacts, where nodes represent individuals and edges represent contact relationships between these individuals (Newman 2010). Understanding the properties of contact networks helps us gain insights into how communicable diseases disperse through a population (Eames and Keeling 2003; Keeling and Eames 2005; Newman 2002; Smith 2006).

Disease dispersion is inherently a spatial process (Bian 2013; Bian et al. 2012). A contact network, once projected into space, becomes a spatially embedded network where nodes are projected according to, for example, individuals' home and workplace locations and edges are projected according to the contact relationship between individuals. The spatial characteristics of disease dispersion can be readily studied in such networks (Zhong and Bian 2016).

Disease dispersion is inherently a spatial process, while scale is involved in all spatial phenomena. Spatial resolution and spatial extent are two common connotations of spatial scale. Spatial resolution is the size of the finest distinguishable areal grains that collectively constitute a study area. It represents the level of detail that is of interest to researchers. Spatial extent is the size of a study area that consists of a large number of areal units (Bian and Walsh 1993; Lam and Quattrochi 1992; Turner et al. 1989). It represents the spatial context of an investigation.

The effect of spatial resolution commonly refers to changes in phenomena properties when areal units are aggregated to different levels, while keeping the same study area. A typical example is the well-known 'modifiable areal unit problem' (MAUP) (Fotheringham 1989; Jiang and Sui 2014; Liu et al. 2014; Openshaw 1983; Openshaw and Taylor 1979). In comparison, the effect of spatial extent refers to changes in phenomena properties in response to enlarged study areas, while keeping the same resolution (Bian and Walsh 1993; Lam and Quattrochi 1992). Many studies are based on an arbitrarily selected spatial extent, and results may not be generalizable to studies of different extents (Turner et al. 2001; Wu and Wu 2013). Between the two connotations, the effect of spatial extent is less studied, and collectively, there are rarely studies addressing the scale effects on network properties.

Network structure is the most important network property, as it determines how nodes are connected and affects the dynamics of epidemics (Eubank et al. 2004; Keeling and Eames 2005; Newman 2010; Smith 2006). Spatially embedded contact networks have two sets of structures, the network structure and the spatial structure. The network structure has been actively studied, while the spatial structure of contact networks has received attention only in recent years. Little is known whether the two structures respond to each other and whether using one could infer the behavior of the other (Barthélemy 2011; Bian 2013; Riley 2007; Tang and Bennett 2010).

Further, the networks are known for their resistance in structures when a fraction of nodes or edges are removed (Albert et al. 2000; Buldyrev et al. 2010; Callaway et al. 2000; Gao et al. 2014; Liu et al. 2011). Most resistance studies, however, have focused on simulated random networks. Results may not be applicable to complex yet common structures in empirically observed networks (Holme 2004; Holme et al. 2002). Empirical network studies, on the other hand, indeed focus on actual networks, but seldom on their resistance properties (Karrer and Newman 2010; Newman 2009). Neither kind of study has looked into the network resistance to spatial structures.

This study aims to examine the scale effects, in terms of changing spatial extent, on the network structure and the spatial structure of contact networks. Specifically, we evaluate (1) the changes in the two contact network structures in response to changes in scale, and (2) the ranges of scale at which contact networks are scale dependent. To achieve these goals, three networks, one observed and two randomly structured, are partitioned into multiple levels of 'unit' networks, each in a smaller, independent spatial extent. The network structure and the spatial structure of the unit networks are compared across scales, where the two structures are represented by six network indices. Two sets of areal units, one set of regular grid and one set of irregularly shaped census unit are used to support the intended scale study.

Findings of this study provide a better understanding of the properties of contact networks at multiple scales. This knowledge could help researchers and policy makers design scale-adaptive strategies to control and prevent communicable diseases effectively.

The remainder of this article is organized as follows. Due to the number of concepts involved in the subsequent discussion, the following background section describes the network structure and the spatial structure, along with the six network metrics. Section 3 introduces the observed contact network data. Section 4 describes the three networks, the two sets of areal units, and the division of networks into unit networks at multiple scales. Section 5 evaluates the scale effects on the networks, and Section 6 summarizes the findings.

## 2 Background

The network structure and the spatial structure of networks refer to how nodes are connected from the network and spatial perspectives, respectively. Component size, clustering coefficient, and average path length are the essential set of metrics used to describe the structure for various networks, including spatially embedded contact networks (Albert et al. 2000; Kovacs and Barabasi 2015; Liu et al. 2011; Newman 2010; Watts and Strogatz 1998). Two additional metrics are considered in this study to measure the spatial structure, the statistical distribution of edge distance and the statistical distribution of the distance of the lost edges when dividing networks into smaller area. Each metric is described below.

Component is a cluster of nodes within a network. All nodes within a cluster are directly or indirectly (through a chain of other nodes) connected to all other nodes within the cluster, but disconnected with nodes in other clusters (Newman 2010). A network can have multiple components. The number of nodes in a component defines its size. The component is a global measurement of how cohesively a network is connected. Two metrics are commonly used to express component size, the relative size of the largest component (denoted as *S*) and the average size of other components (denoted as <*s*>) (Newman 2010; Wasserman and Faust 1994). The relative size of the largest component is the ratio of the size of the largest component to the size of the network:

$$S = \frac{n_{max}}{n} \tag{1}$$

where $n_{max}$ is the size of the largest component, and *n* is the size of the network (the total number of nodes in the network). The average size of other components is defined as:

$$\langle s \rangle = \frac{\sum_1^c s_i}{c-1} \qquad i \neq max \tag{2}$$

where $s_i$ is the size of component $i$, and $c$ is the total number of components in the network. A greater $S$ value indicates a more cohesive network, while a smaller $<s>$ value also indicates the same. For cohesive networks, a large $S$ value usually accompanies a small $<s>$ value. Otherwise, for fragmented networks, both $S$ and $<s>$ values can be low.

The clustering coefficient of a node is the number of connections between its direct neighboring nodes, divided by the number of all possible connections between these nodes. This metric represents local clustering by measuring how tightly a node's neighbors are clustered together (Watts and Strogatz 1998). Equation 3 expresses the clustering coefficient $c_i$ of node $i$ as:

$$c_i = \frac{2e_i}{k_i(k_i-1)} \qquad (3)$$

where $k_i$ is the number of neighboring nodes of $i$, and $e_i$ is the number of connections between the neighboring nodes. The clustering coefficient of an entire network is the average over the clustering coefficients of all nodes:

$$cc = \frac{\sum_{i=1}^{n} c_i}{n} \qquad (4)$$

A higher $cc$ means a stronger locally clustered structure. Within a component, there may exist a number of highly localized clusters.

The path length is the number of consecutive edges between a pair of nodes. Among all possible paths between the two nodes, the one with the shortest length is called the shortest path. The average path length of the entire network is the average of the shortest paths between all possible pairs of nodes (Watts and Strogatz 1998). This metric measures the efficiency of how a node can be connected from any other node in the component. It is defined as:

$$l = \frac{1}{n(n-1)} \sum_{i \neq j} l(v_i, v_j) \qquad (5)$$

where $l(v_i, v_j)$ is the length of the shortest path between nodes $v_i$ and $v_j$. A shorter $l$ implies a more efficiently connected network structure. As an absolute measurement, this metric is sensitive to network size when the network is divided into multiple levels of smaller size. To eliminate this effect and be consistent with the relative scale of $S$ and $c$, $l$ is standardized as the relative average path length $l'$:

$$l' = \frac{l}{lmax} \qquad (6)$$

where $lmax$ is the diameter of the network, i.e. the maximum of all shortest path length in a network (Watts and Strogatz 1998).

The statistical distribution of edge distance (*Dist*) in a network measures the spatial structure (Barthélemy 2011). A negatively skewed distribution indicates the dominance of short edges, thus a spatially clustered structure, while a positively skewed distribution implies a spatially sparse network. Otherwise a normal distribution indicates a spatially random network. When the original network is divided into multiple levels of unit networks in smaller spatial extent, those edges that extend across boundaries of the spatial extent are eliminated, while those within the spatial extent are preserved. In this sense, *Dist* is the distribution of remaining edges at each scale. The response of *Dist* to scale change may indicate whether the spatial structure is resistant to edge removal or not. The statistical distribution of distance of the lost edges (*Loss*), the second spatial metric, is considered the complement to *Dist* because it is the difference in *Dist* between the original network and the network at a specific scale. The six indices *S*, *<s>*, *cc*, *l'*, *Dist*, and *Loss*, are used to analyze the structure of contact networks.

**3 Contact network**

The contact network used in this study was constructed previously by Bian et al. (2012) for a residential area in a metropolitan community in the Northeastern US. This area covers 495 census blocks, 72 census block groups, and 22 census tracts in an area of approximately 4800 m * 3700 m. The network consists of 64,726 individuals. Each individual is assigned to a family and most individuals are also assigned to a workplace (including schools). The individuals, households, workplaces (including schools), and society-wide network were constructed using a wide range of data, such as demographic, social-economic, occupational, commuting, income, vehicle ownership, workplace type, and spatial distributions of households and workplaces.

The constructed households confirmed census statistics. The constructed co-workers confirmed multiple sets of public information, such as census statistics, a regional household survey, and a regional commuting survey (Bian et al. 2012). Simulated influenza epidemics using this contact network showed a good agreement with CDC's weekly report for an influenza epidemic for the study area (Bian et al. 2012).

There are two types of contact relationships between individuals. One type is the contact between family members and another is between co-workers. The network represents individuals as nodes and the contact relationships between them as edges, resulting in a total of 64,726 nodes and 194,683 edges. The two types of relationships are treated as family edges (93,474) and co-worker edges (101,209), respectively. To examine the scale effects, the contact network is projected into space. Nodes are projected according to their home locations, and all nodes that represent members of a family share an identical location (Fig. 1). Edges are projected into space according to the location of their associated nodes. The distance of the family edges is zero. A co-worker edge links two individuals who work in the same workplace but may reside at two different home locations. These co-worker edges vary in distance. The properties of this network, called an observed network, along with two other networks, are discussed in the following text.

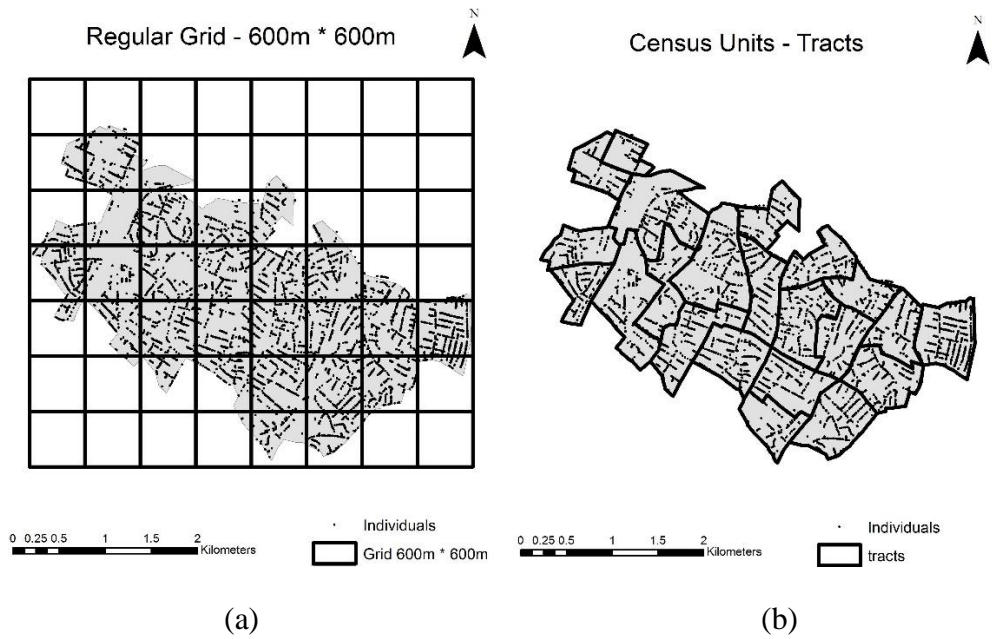

(a)                  (b)

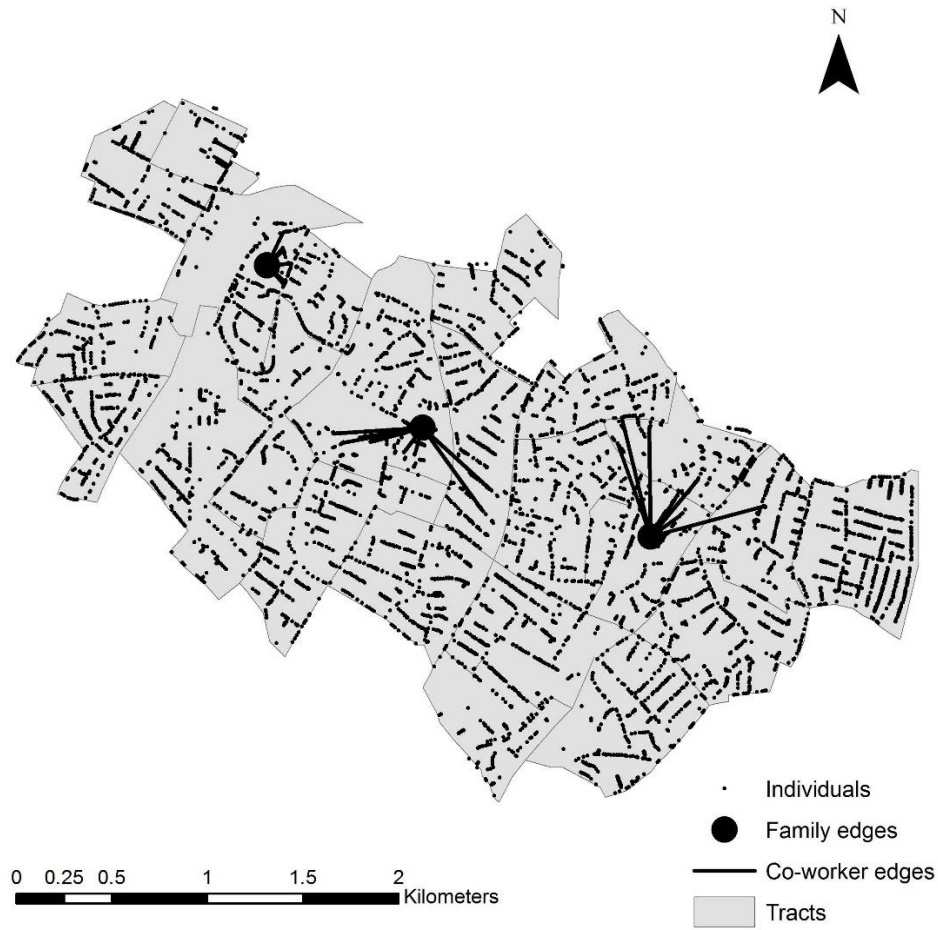

(c)

(d)

**Fig. 1.** The study area and its division (boldface lines). Dots are locations of individual nodes. (a) The study area divided by the regular grids, (b) the study area divided by census units, (c) co-worker edges of three different families, and (d) an illustration of family edges (boldface) and co-worker edges. The 0-distance family edges are intentionally exaggerated for illustration clarity.

**4 Methods**

In order to investigate the scale effects on the structures of the networks, this study divides networks into multiple scales and examines if and how their properties change with scales. The three networks are the observed network as discussed above, a random-node network, and a random-edge network. Two sets of areal units, regular grids and census units, are applied to divide the three networks into multiple scales. The six network structure metrics are analyzed and compared between the three networks and between the two sets of areal units. The three networks are discussed first below, followed by the discussion of the two sets of areal units, and then the division of networks.

4.1 Three networks

Two random networks, a random-node network and a random-edge network, are generated to systematically examine the network structure and the spatial structure of contact networks. As properties of random networks are controllable, they are commonly used as references to study behaviors of observed networks, and research findings can then be extended to a broad range of networks (Latapy et al. 2008; Liu et al. 2011; Luo et al. 2014; Ruths and Ruths 2013; Salathé and Jones 2010). The two random networks are designed to have the same basic properties as the observed network in terms of three constraints. The three networks use an identical number of nodes, an identical number of edges, and an identical distribution of 'degree' of the observed network. The degree is the number of neighbors of a node. The statistical distribution of node degree is a basic constraint of the network structure (Freeman 2004; Newman 2010; Wasserman and Faust 1994).

The random-node and random-edge networks preserve the network structure and the spatial structure of the observed networks, respectively, while altering the other (Fig. 2). Specifically, the random-node network alters the spatial structure of the observed network by randomizing node locations. The associated edge location and the statistical distribution of edge distance change according to the new locations of nodes. Yet, the random-node network maintains an identical network structure as the observed network in terms of how nodes are connected (Fig. 2b). In contrast, the random-edge network alters the network structure by randomly shuffling edges between nodes, while keeping the locations of nodes. The random-edge network maintains an identical spatial structure as the observed network by following the same statistical distribution of edge distance (Fig. 2c).

Current network randomization algorithms are only concerned with the network structure, not the spatial structure (Barthélemy 2011; Newman 2010). This study devises a spatially explicit randomization algorithm to address the joint probability distribution of the network structure and the spatial structure for the two random networks (Fig. 2). Each of the two random networks is generated 1,000 times, and the six metrics are calculated for each simulation of the two networks. The average values of the metrics are used to represent the properties of the two random networks for the subsequent analysis. The summary statistics in Table 1 illustrate the properties of the three networks before they are divided (*Loss* is not reported here, because it is only available after the networks are divided). The shared and distinctive network structure and spatial structure properties among the three networks may subject them to different scale effects.

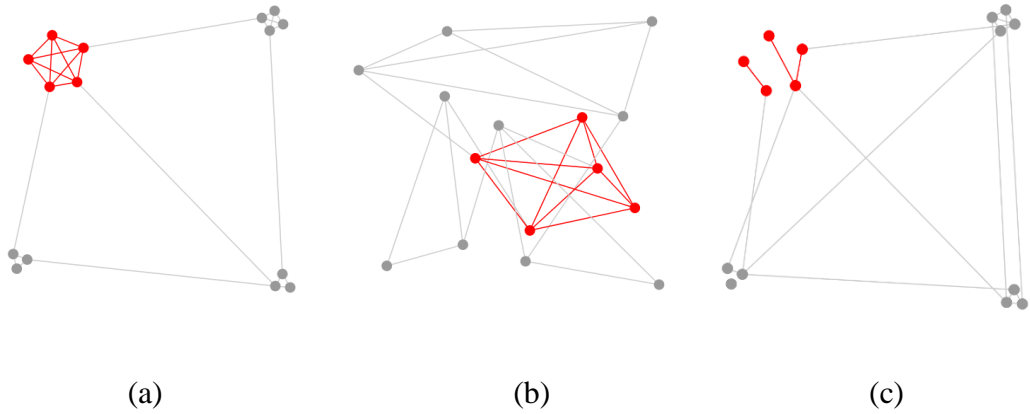

(a)              (b)              (c)

**Fig. 2.** An illustration of the observed network (a), the random-node network (b), and the random-edge network (c). To illustrate the 0-distance family edges (red edges in 2a), the physical distance between family members are intentionally exaggerated.

**Table 1** Summary statistics of the three networks.

|  |  | Observed network | Random-node network (average) | Random-edge network (average) |
|---|---|---|---|---|
| Basic properties | Number of nodes | 64,726 | 64,726 | 64,726 |
|  | Number of edges | 194,683 | 194,683 | 194,683 |
|  | Average degree | 6.01 | 6.01 | 6.01 |
| Network structure | Relative size of the largest component (%) | 83.70 | 83.70 | 92.37 |
|  | Average size of other components | 1.51 | 1.51 | 1.02 |
|  | Clustering coefficient | 0.43 | 0.43 | 0.08 |
|  | Relative average path length | 0.30 | 0.30 | 0.39 |
| Spatial Structure | Average edge distance (m) | 327.11 | 1687.02 | 327.11 |

4.2 Two sets of areal units

Two sets of areal units, regular grids and census units, are used to divide the networks. Using the regular grids, the study area is divided into a total of 24 levels of regular grids ranging from 100 m * 100m to 2400 m * 2400 m using 100 m increments. The minimum cell size is comparable to the size of a census block. The maximum cell size is the largest possible area that can be used to divide the study area into multiple cells. Each level represents a scale of spatial extent. Because the outer boundary of the regular grids is not consistent with that of the study area, the grids include both empty cells and cells that partially overlap with the study area. The empty cells are discarded. For the partial cells, the metric values are computed according to the proportion of the cell that falls within the study area.

The census units at three scales, including blocks, block groups, and tracts, are employed because they are a well-established means to organize many socio-economic and demographic data. In addition, they are spatially explicit and available at multiple scales. The average size of blocks, block groups, and tracts are equivalent to the 100 m * 100 m, 300 m * 300 m, and 600 m * 600 m cells, respectively.

4.3 Division of networks

During the division, those edges that extend across boundaries of spatial extent are eliminated, while those within the areal units are kept. This results in a greater number of smaller-sized unit networks at each finer scale. Each unit network is an independent network in an isolated study area of smaller spatial extent. Each unit network has its own network structure and spatial structure. The values for the six metrics ($S$, $<s>$, $cc$, $l'$, $Dist$, and $Loss$) are calculated for each unit network. Changes in the properties of unit networks across scales of spatial extent are evaluated first for each network, and then compared between the three networks and between the two sets of areal units.

**5 Results and discussion**

Given the two sets of areal units and the three networks, the properties of unit networks are analyzed for the following six (2 * 3 = 6) combinations of areal units and networks: (1) regular grids combined with the observed network, the random-node network, and the random-edge network, and (2) census units combined with the observed network, the random-node network, and the random-edge network. The three combinations associated with the regular grids are discussed first, followed by the three combinations with the census units.

5.1 Regular grids

5.1.1 Regular grids + observed network

The values of the four network structure metrics, i.e. $S$, $<s>$, $cc$, and $l'$, and the two spatial structure metrics, i.e. *Dist* and *Loss*, for all unit networks are plotted against the 24 cell sizes in Fig. 3. Because there are 24 sets of edge distance distributions, one for each grid size, for illustration clarity, *Dist* and *Loss* for cell sizes of 600 m * 600 m, 1200 m * 1200 m, and 2400 m * 2400 m, are shown in Fig. 3e-f. Fig. 3(e) also includes *Dist* of the three networks before they are divided.

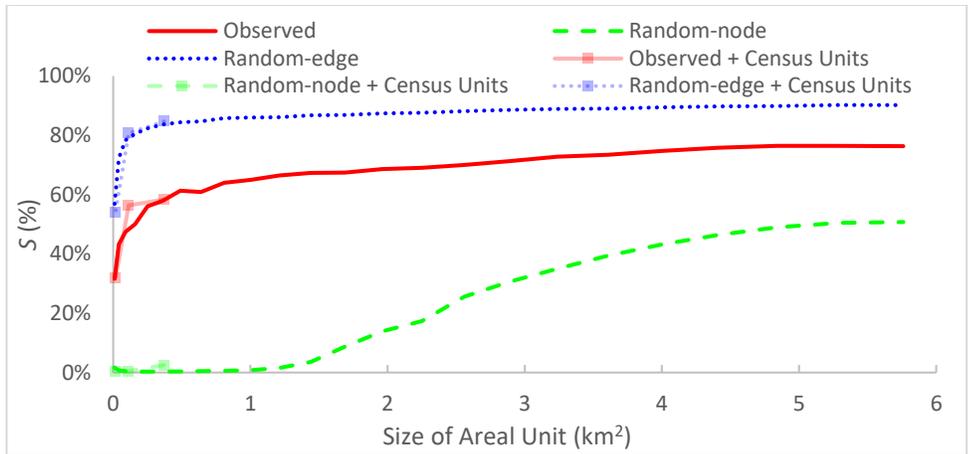

(a)

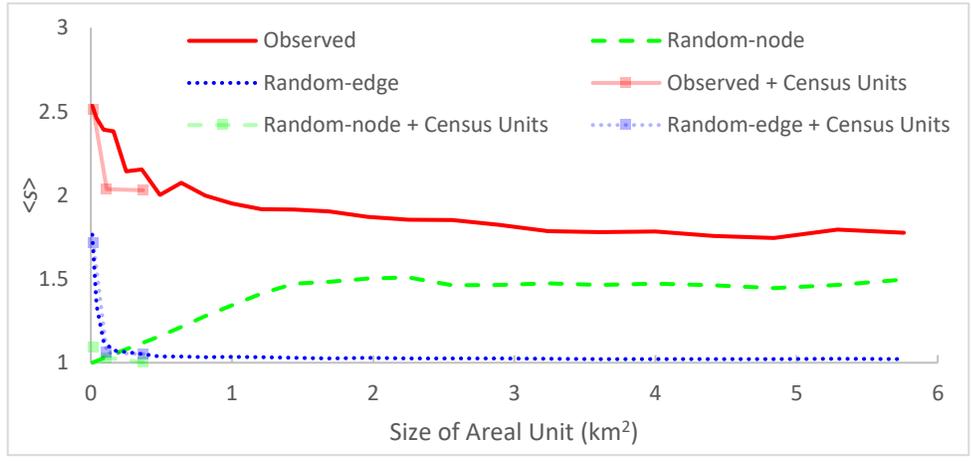

(b)

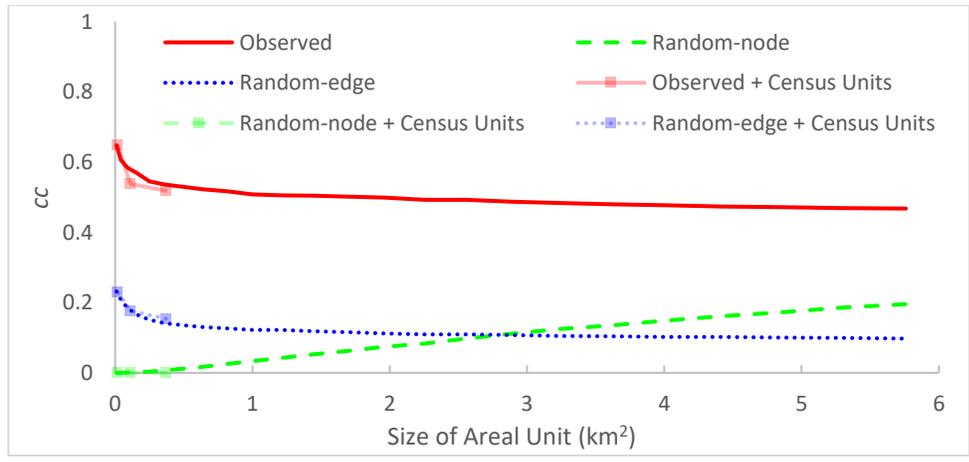

(c)

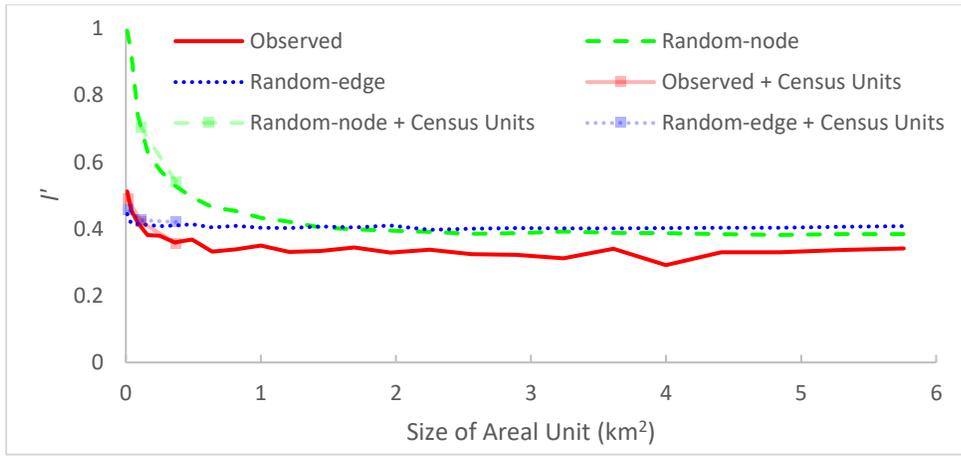

(d)

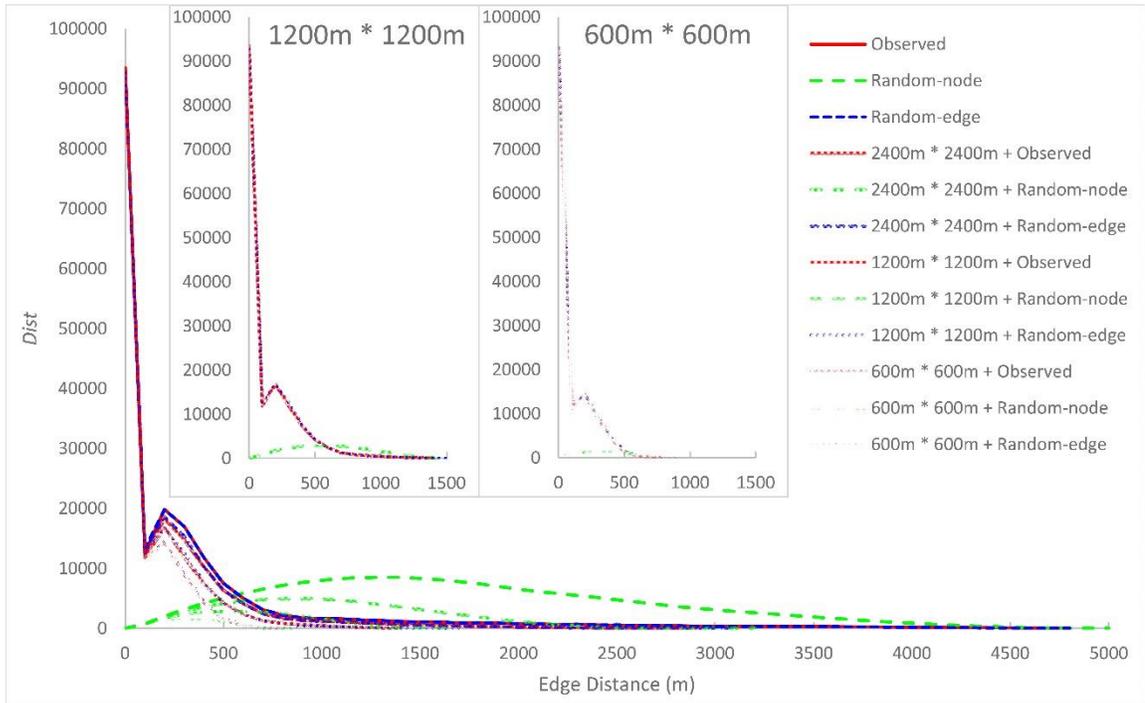

(e)

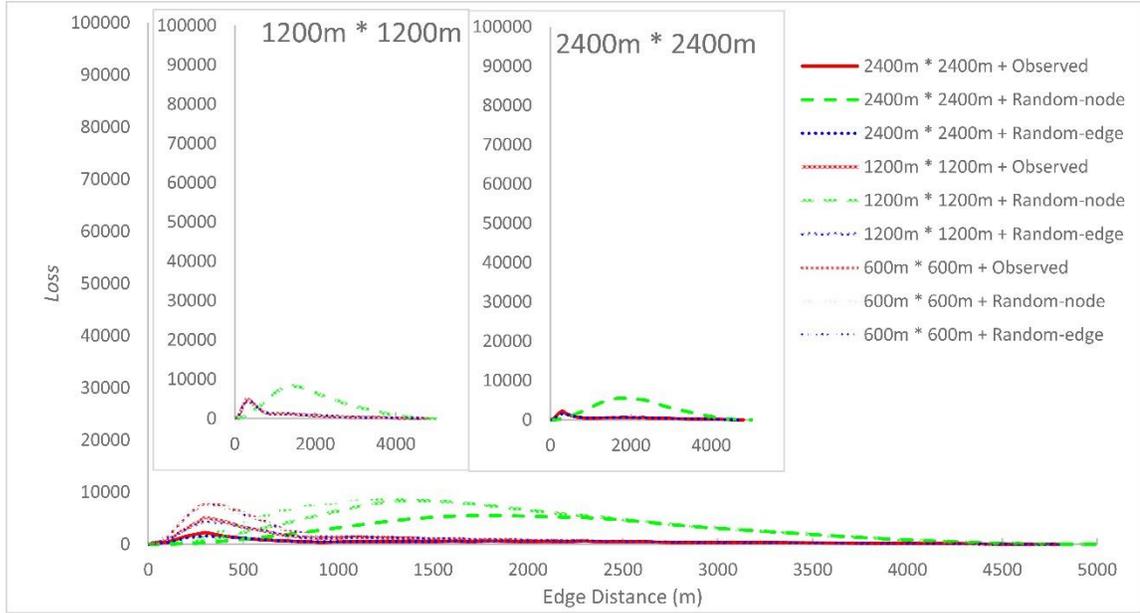

(f)

**Fig. 3.** (a) The relative size of the largest component $S$ (%), (b) the average size of other components $<s>$, (c) the clustering coefficient $cc$, (d) the relative average path length $l'$, (e) the statistical distribution of edge distance $Dist$ (including that of the three original networks), and (f) the statistical distribution of the lost edge distance $Loss$. Fig. 3a-d also show $S$, $<s>$, $cc$, and $l'$, respectively, averaged over census blocks, block groups, and tracts.

The $S$, $<s>$, $cc$, and $l'$ values of the observed network (red lines, Fig. 3a-d) vary with scale, showing a general trend of scale dependence. The $S$ curve rises rapidly up to a characteristic scale of 0.6 km$^2$, and begins to level off afterwards. On the other hand, the $<s>$, $cc$, and $l'$ values decrease with scale, and their variations correspond reversely to the $S$ curve (Fig. 3a-d). The metric values collectively indicate that the unit networks at fine scales are globally fragmented (low $S$), locally clustered (high $cc$), inefficient (high $l'$), and consequently robust against disease dispersion. Beyond 0.6 km$^2$, all four metrics stabilize, behaving independently of scale change. The network structure at these coarser scales is more cohesive (higher *S*), remains clustered (similar *cc*), and is more efficient (lower *l'*), thus are more vulnerable to disease dispersion.

*Dist* appears to be independent of scale. Although absolute quantities of edge distance change (a lesser number of edges and shorter distances at finer scales), the response of *Dist* to the scale change is invariant. This suggests that the spatial structure of the unit networks is resistant to edge loss across scales. The *Dist* distribution shows a strong distance-decay pattern. It peaks at 0 m and diminishes at approximately 800 m or shorter. The peak reflects the large number of 0-distance family edges, which are not affected when the network is divided into smaller spatial extents. The second peak between 0 and 800 m is caused by a large supply of short co-worker edges. The 800 m diminishing point is equivalent to the characteristic scale of 0.6 km$^2$. Such a short-edge dominant pattern indicates that the unit networks are highly clustered in space at all scales. Spatially clustered networks facilitate short distance disease dispersion and lead to an epidemic surge in small areas.

*Loss*, as a complement to *Dist*, is the difference in edge distance distribution between the original network and the unit networks at a given scale (Fig. 3f). Although absolute quantities of the lost edge distance change with scale, the response of *Loss* to scale change persists through scales. *Loss* provides supplementary evidence to the behavior observed in *Dist*, i.e. scale independence, spatial clustering, and resistance to edge loss. The two metrics imply that, unlike the network structure that is scale dependent in a scale range (0.01-0.6 km$^2$), the spatial structure of the unit networks are independent of scale changes.

5.1.2 Regular grids + random-node network

For the random-node network (green lines, Fig. 3a-d), its network structure is also scale dependent, but in a different manner from that of the observed network. At the fine scale range of 0-1.4 km$^2$, the unit networks are extremely fragmented ($S$ close to 0 and low $<s>$ value), have barely any cluster ($cc$ value close to 0), and are extremely inefficient ($l'$ value close to 1). In contrast, beyond the characteristic scale of 1.4km$^2$, the unit networks are more cohesive (higher $S$ and $<s>$ values), clustered (higher $cc$ value), and efficient (lower $l'$ value). Collectively, the behavior of the random-node network gradually approaches that of the observed network at coarse scales.

*Dist* of the random-node network appears to be independent of scale. While absolute quantities of edge distance change, its negatively skewed normal distribution persists across all 24 scales (4 are shown in Fig. 3e), showing resistance to edge loss. Its peak at 1200 m corresponds to the 1.4km$^2$ characteristic scale. The randomization of node locations lengthens the edge distance randomly (Table 1), shown as a normal distribution (Fig. 3e). The random-node network breaks away from the spatially clustered structure in the observed network and is a spatially scattered network. *Loss* offers supplementary evidence to the unit network properties as observed in *Dist* behavior. In contrast to the network structure of the random-node network that is scale independent, the spatial structure is independent of scale changes.

5.1.3 Regular grids + random-edge network

For the random-edge network (blue lines, Fig. 3a-d), its network structure, represented by $S$, $<s>$, $cc$, and $l'$ curves, are in parallel to those of the observed network, but considerably deviated from each other. The characteristic scale is much finer than that of the observed network. The unit networks of the random-edge network are much more cohesive (much higher $S$ and much lower $<s>$), much less clustered (much lower $cc$), and mostly less efficient (lower $l'$) (Table 1).

In terms of spatial structure metrics, *Dist* and *Loss*, the behavior of the random-edge network is almost identical to that of the observed network. The spatial structure seems to be independent of scale and resistant to the edge removal.

5.2 Census units

Fig. 4a-d show the four network structure metrics, $S$, $<s>$, $cc$, and $l'$, of all unit networks plotted against the size of three types of census units (red symbols in different shape and shade). The two spatial structure metrics, *Dist* and *Loss*, at each census scale are shown in Fig. 4e-f, respectively (red solid lines in three shades).

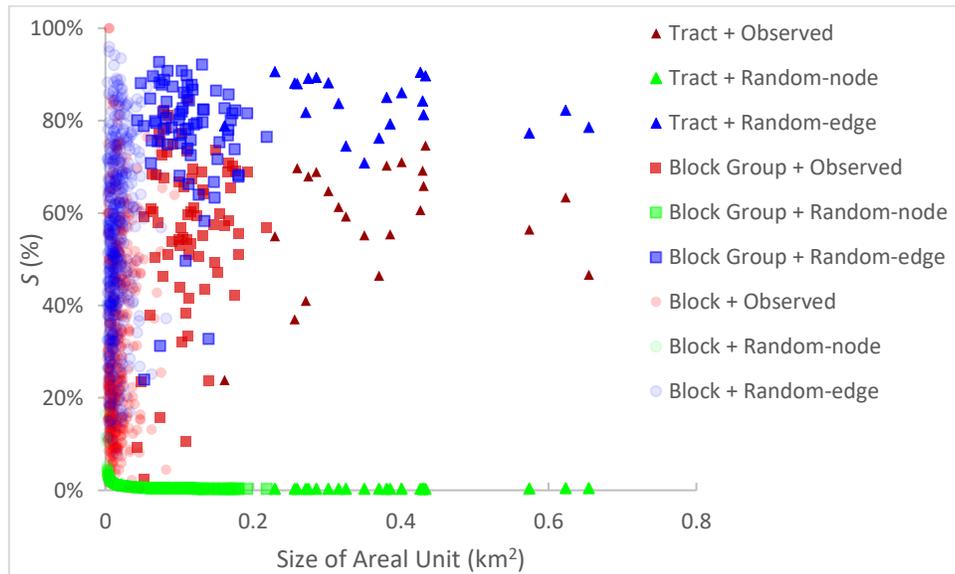

(a)

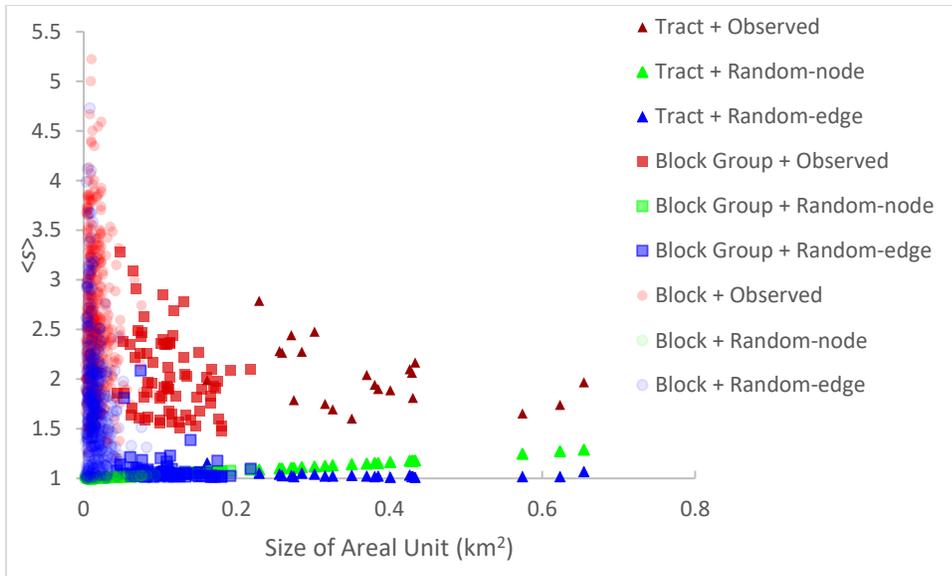

(b)

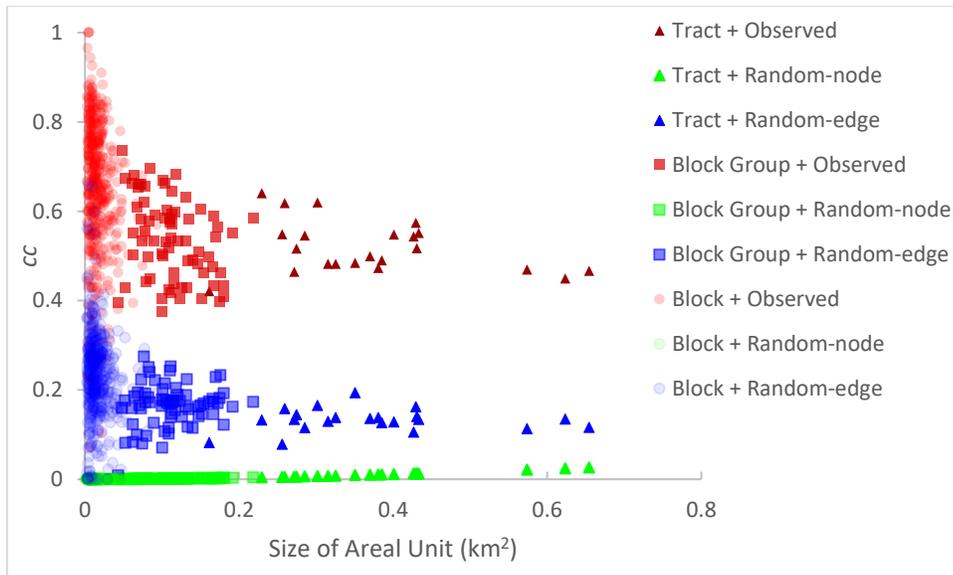

(c)

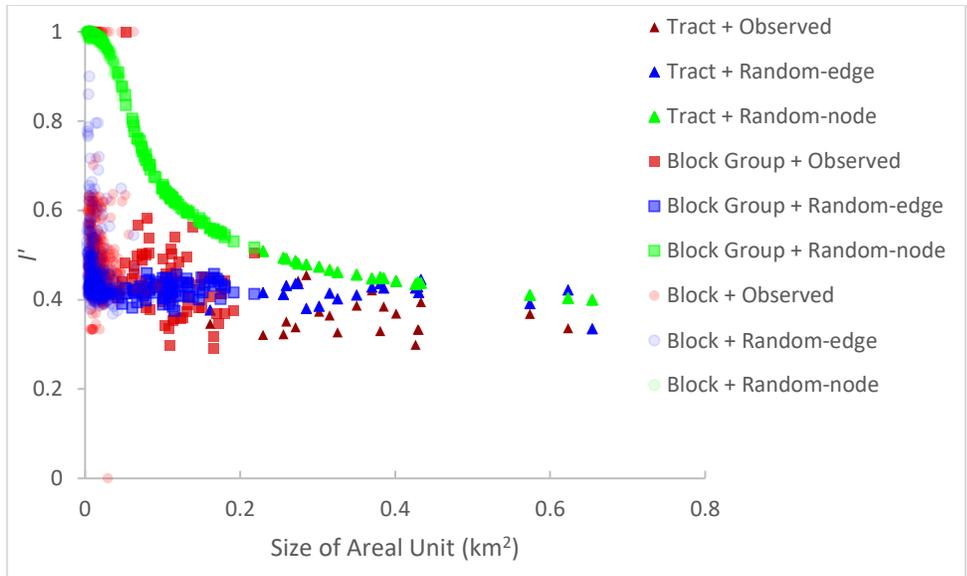

(d)

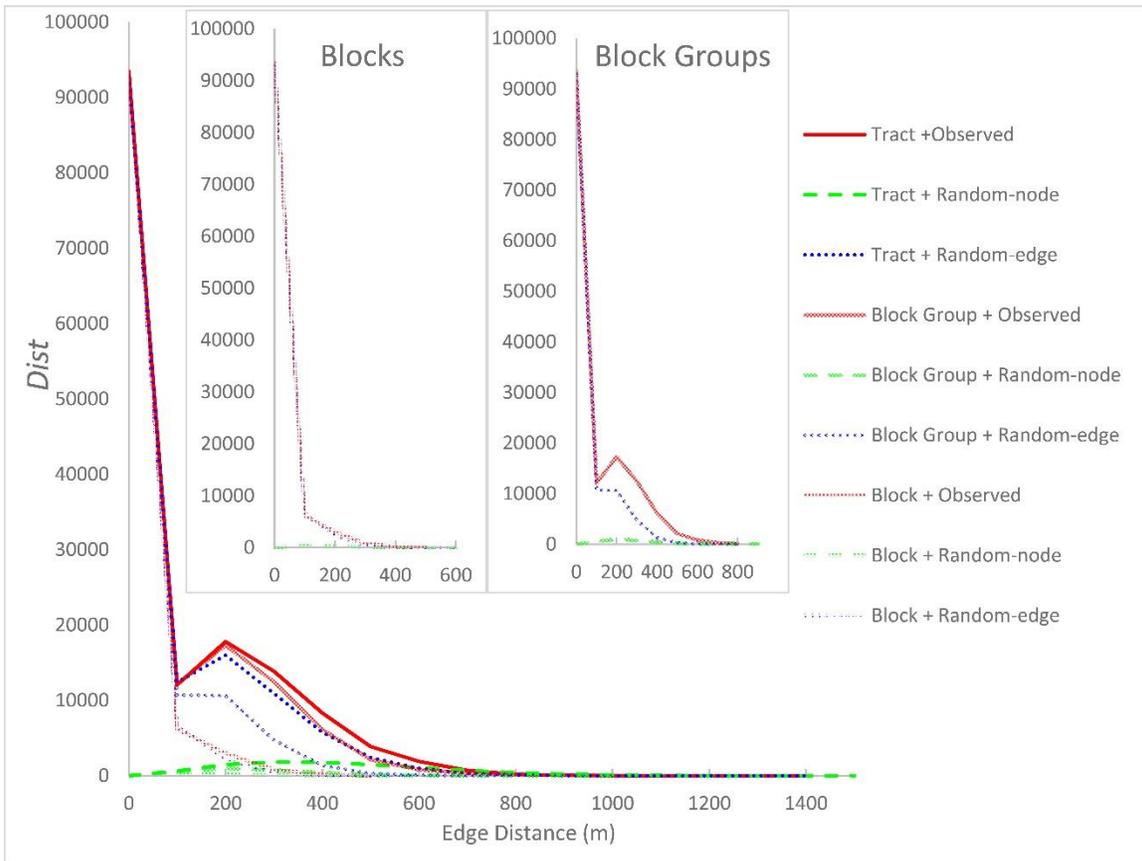

(e)

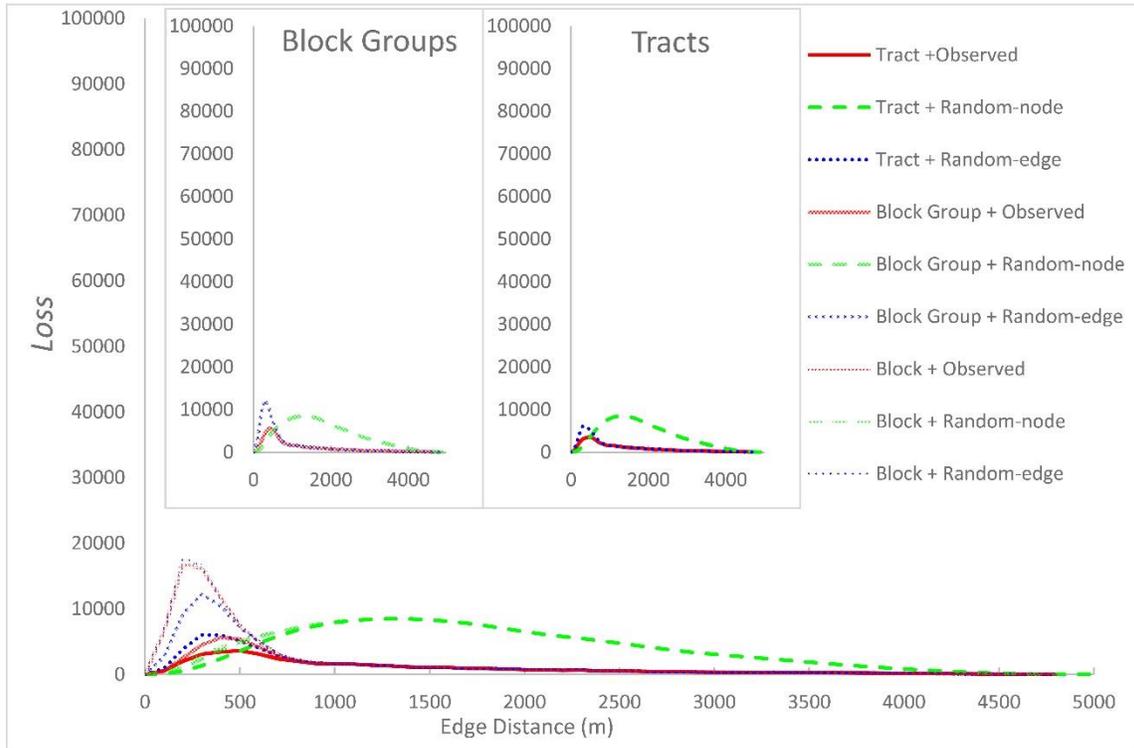

(f)

**Fig. 4.** (a) The relative size of the largest component *S*, (b) the average size of other components <*s*>, (c) the clustering coefficient *cc*, (d) the relative average path length *l'*, (e) the statistical distribution of edge distance *Dist*, and (f) the statistical distribution of the lost edge distance *Loss* for all three networks using census units. The two inserts in (e) show the enlarged *Dist* for block groups and blocks at their short edge distance portion. The two inserts in (f) are the enlarged *Loss* for block groups and tracks at their short edge distance portion.

The network structure metrics of all three networks show scale dependence across census scales, in terms of $S$, $<s>$, $cc$, and $l'$ (Fig. 4a-d). This trend coincides with the scale dependence for regular grids at the corresponding scale range, mostly at fine scales (Fig. 3a-d). The two spatial structure metrics, *Dist* and *Loss,* on the other hand, are independent of scale. The spatial structure of the three networks, either clustered or scattered, is resistant to edge loss (Fig. 4e-f). This trend is also consistent with that of the regular grids (Fig. 3e-f). The behaviors of all three networks show good agreement between the regular grids and the census units as far as the network structure and the spatial structure are concerned. It seems that scale has inherent effects on networks regardless of the shape of areal units used to divide the networks. The behaviors of all three networks show good agreement between the regular grids and the census units as far as the network structure and the spatial structure are concerned. It seems that scale has inherent effects on networks regardless of the shape of areal units used to divide the networks.

5.3 Discussion

The two random networks as references help reveal the unique properties of the observed network. When comparing to the reference network with randomized network structure (random-edge network), the observed network tends to be 'dense', thus cohesive, clustered, and efficient, and is sensitive to scale change at fine scales (Section 5.1). When comparing to the reference network with randomized spatial structure (random-node network), the observed network is spatially clustered, and the clustering pattern is resistant to edge loss, thus it is independent of scale (Fig. 3e).

The two structures are closely related. Spatially, the co-worker edges are mostly short (< 800m), in addition to the 0-distance family edges (Fig. 3e). This can be attributed to urban dwellers' preference of being in close proximity to schools and workplaces (Ben-Akiva and Lerman 1985). Those who live close to each other tend to go to the same schools or workplaces. In other words, the closer ones are more connected.

These short distance co-worker edges are mostly affected when the network is divided into fine scales where the co-worker edges begin to be eliminated (Fig. 3e-f). This 800 m or smaller clustering spatial structure may affect the behavior of the network structure. At the characteristic scale at 800 m or above, the behavior of the network structure stabilizes (Fig. 3). Here, the 800 m is considered the 'operational scale' where the fundamental process (co-worker connection) operates at (Bellier et al. 2007; Bian and Walsh 1993; Fortin et al. 2012). Scale effects of networks cannot be studied on their network structure alone without considering their spatial structure.

However, the spatial structure alone is not sufficient to infer the network structure. The observed network and the random-edge network share an identical spatial structure at the original scale (Table 1), but their network structures are different at all scales (Fig. 4a-d). For example, in terms of how nodes are connected, all family nodes (or co-worker nodes) are directly connected to all other nodes within a family (or a workplace), resulting in many redundant edges in the observed network. Taking a family of five members as an example, the observed network requires a total of $n(n-1)/2$ edges, i.e. 10 edges. A more concise network structure requires as few as $(n-1)$, or 4 edges. The random-edge network randomly redistributes the redundant family and co-worker edges to connect a greater number of other nodes into a greater sized largest component than that of the observed network (Table 1).

Similarly, the network structure alone is not sufficient to infer the spatial structure. The observed network and the random-node network possess an identical network structure, but their spatial structure differs. The observed network is highly clustered in space, in contrast to the scattered spatial structure of the random-node network (Fig. 3e). A comprehensive understanding of the scale effects requires examination of both structures.

The property of 'the closer ones are more connected' discussed above represents the network structure, spatial structure, and the relationship between the two structures for contact networks. This property is distinct, but not unique to contact networks. Many other spatially embedded networks, such as human mobility networks, mobile phone networks, social media networks, and friendship networks, show distance-decay effects similar to the observed network discussed in this study, although to different degrees (Brockmann et al. 2006; Crandall et al. 2010; Eagle et al. 2009; González et al. 2008; Jiang et al. 2009; Liu et al. 2012; Liu et al. 2014). The findings about the dual structure and the relationship between them may be generalized to these networks, among many others.

The cohesive, clustered, and efficient network structure and highly clustered spatial structure of contact networks are vulnerable to disease dispersion. Yet, these properties carry important implications in the design of scale-adaptive strategies to control and prevent the dispersion of communicable diseases, such as multi-scale household and community quarantine strategies (Bajardi et al. 2011; Camitz and Liljeros 2006; Epstein et al. 2007; Ferguson et al. 2006; Longini et al. 2005; Mao 2013). The characteristic scale, such as 0.6km$^2$ (800m), may help determine the scope of quarantine.

## 6 Conclusions

This study examines the scale effects, in terms of spatial extent, on the network structure and the spatial structure of spatially embedded contact networks. Two issues are explored, how the two types of structure change in response to scale change, and the range of scale effects. In terms of the network structure, the properties of the observed network are sensitive to scale changes at fine scales. In comparison, the clustered spatial structure of the observed network is scale independent.

Results of this study inform the user of the selection of an appropriate scale for network studies. Both the network structure and spatial structure are vital in this selection. Further, this study establishes the relationship between the network structure and the spatial structure of contact networks. The behaviors of the network structure are affected by the spatial structure, but either one is insufficient to infer the other. The analysis of spatial structure of contact networks provides valuable information about the spatial pathways of disease dispersion and the affected areas. This information helps identify households and communities that are vulnerable to health risks and helps deploy intervention strategies to spatially targeted areas.

Zhong, S., & Bian, L. (2016). A location-centric network approach to analyzing epidemic dynamics. *Annals of the American Association of Geographers, 106*, 480-488